\documentclass[journal]{IEEEtran}

\usepackage{array}
\newcolumntype{C}[1]{>{\centering\arraybackslash}m{#1}}
\newcolumntype{N}{@{}m{0pt}@{}}

\usepackage{cite}
\usepackage[symbol]{footmisc}
\usepackage{hyperref}
\usepackage{csquotes}
\usepackage{balance}
\usepackage{color}
\usepackage{tabularx}
\usepackage{array}
\usepackage{caption}
\usepackage{subcaption}
\usepackage{graphicx}
\usepackage{multirow}
\usepackage{amsmath}
\usepackage{lipsum}
\usepackage{enumitem}
\usepackage{makecell}
\usepackage[table]{xcolor}
\usepackage{rotating}
\usepackage{amssymb}
\usepackage{pifont}
\usepackage{booktabs,caption}
\usepackage[flushleft]{threeparttable}

\newcommand{\cmark}{\ding{51}}%
\newcommand{\xmark}{\ding{55}}%

\begin{document}

\title{Internet of Everything Enabled Solution for COVID-19, its new variants, and Future Pandemics: Framework, Challenges, and Research Directions}	

\author{Sunder Ali Khowaja, Parus Khuwaja, and Kapal Dev*, \IEEEmembership{Member IEEE}

\thanks{*Corresponding and senior author}

\thanks{Sunder Ali Khowaja is an Assistant Professor with Department of Telecommunication, Faculty of Engineering and Technology, University of Sindh, Pakistan.)}

\thanks{Parus Khuwaja is an Assistant Professor with Institute of Business Administration, University of Sindh, Pakistan.)}

\thanks{Kapal Dev is with the Nimbus Research Centre, Munster Technological University, Ireland e-mail:kapal.dev@ieee.org}
}

\maketitle
\begin{abstract}
After affecting the world in unexpected ways, COVID-19 has started mutating which is evident with the insurgence of its new variants. The governments, hospitals, schools, industries, and humans, in general, are looking for a potential solution in the vaccine which will eventually be available but its timeline for eradicating the virus is yet unknown. Several researchers have encouraged and recommended the use of \textit{good practices }such as physical healthcare monitoring, immunity-boosting, personal hygiene, mental healthcare, and contact tracing for slowing down the spread of the virus. In this article, we propose the use of wearable/mobile sensors integrated with the Internet of Everything to cover the spectrum of good practices in an automated manner. We present hypothetical frameworks for each of the good practice modules and propose the COvid-19 Resistance Framework using the Internet of Everything (CORFIE) to tie all the individual modules in a unified architecture. We envision that CORFIE would be influential in assisting people with the \textit{new normal} for current and future pandemics as well as instrumental in halting the economic losses, respectively. We also provide potential challenges and their probable solutions in compliance with the proposed CORFIE.
	\end{abstract}
	
	\begin{IEEEkeywords}
Internet of Everything, IoT Architectures, Wearable Sensors, COVID-19, Emerging Technologies
	\end{IEEEkeywords}
	
	%
	\IEEEpeerreviewmaketitle

\section{Introduction} \label{Introduction}
	
\IEEEPARstart{T}{he} novel coronavirus (COVID-19) has affected the world on a large scale and has compelled people to significantly alter the course of their lifestyles. At the time of this writing, COVID-19 has been transmitted to more than 78.5 million people along with unprecedented deaths, i.e. around 1.7 million \cite{meter}. Multidimensional changes have been observed in people due to COVID-19 situations such as job loss, educational changes, shortage of supplies, interpersonal relationships, loss of loved ones, fear of infection or illness, occupational stress, social isolation, mental health, financial distress, and so forth. Similarly, countries are also facing societal, healthcare, and financial challenges in this pandemic.
	
The novel coronavirus (COVID-19) has affected the world on a large scale and has compelled people to significantly alter the course of their lifestyles. At the time of this writing, COVID-19 has been transmitted to more than 78.5 million people along with unprecedented deaths, i.e. around 1.7 million \cite{meter}. Multidimensional changes have been observed in people due to COVID-19 situations such as job loss, educational changes, shortage of supplies, interpersonal relationships, loss of loved ones, fear of infection or illness, occupational stress, social isolation, mental health, financial distress, and so forth. Similarly, countries are also facing societal, healthcare, and financial challenges in this pandemic. Patients are experiencing a lack of healthcare services due to the shortage of facilities. The government’s ban on gatherings, congregations, and travel restrictions, had a drastic impact on both the country’s economy as well as the financial dynamics of the citizens. Work from the home strategy was chosen as an alternative which made specific occupations archaic, hence, affecting the lifestyle to great lengths.   
The contraction of COVID-19 causes loss of smell and taste, breathing problems, cough, and fever, respectively. The extended length of the aforementioned symptoms in vulnerable patients might result in heart issues, respiratory problems, hypertension, organ failure, and in the worst case death \cite{Dev2020}. In the beginning, older populations were considered to be at risk but as per WHO guidelines “\textit{the children and young people are not invulnerable to this virus}” \cite{Dickinson2020}.  Recently, a COVID-19 variant and mutation, i.e. VUI-202012/01 is under investigation by the Genomics UK consortium \cite{geno}. One thing which is of great concern is that N501Y mutation has been discovered which allows the virus to get bonded with the human ACE2 receptor using spike proteins. This bondage allows the virus to be more infectious and easily spreadable \cite{Wise2020}. Although it’s still in question whether it is deadlier than its former version and the existing vaccine (ChAd0x1 nCoV-19 and BNT162b2) can be effective for the mutated virus or not? Furthermore, the variant was discovered in the UK recently, therefore with utmost certainly the virus can have mutated in different countries as well. According to the study \cite{geno,Wise2020},the Genomics UK consortium has revealed that around 4000 mutations have already been recorded in the spike protein and that more mutations will occur as it is naturally part of the evolution process. It has been observed with the seasonal flu vaccines that they need to be altered every year due to the constant mutations, therefore, it is not wrong to say that the vaccines for COVID-19 and future pandemics need respective adjustments with respect to the mutation which prolongs the pandemic itself \cite{Wise2020,TheLancetMicrobe2020}. Furthermore, it should be noted that the flu vaccine after alteration does not take much time as the base process is available to the manufacturers whereas for Coronavirus the manufacturers are not yet licensed and the base process has not been laid out \cite{col}. Hence, the vaccine is a potential solution but it might take months or years to eliminate COVID-19 provided that either the vaccine effectively handles the variants or the virus stops mutating altogether.
The experts have been repeatedly asked the question regarding the possible timeline for COVID-19 expulsion. Currently, there are a total of 5 vaccines that have entered phase 3 trails BBIBP-CorV (Sinopharm), Sputnik V from Gamaleya Research Institute, ChAd0x1 nCoV-19 from AstraZeneca and University of Oxford, mRNA-1273 from Moderna, and BNT162b2 from the Moderna as well \cite{TheLancetMicrobe2020}. Out of the five, the data for only ChAd0x1 nCoV-19 \cite{voysey2020safety} and BNT162b2 \cite{polack2020safety} has been published and approved by the drug regulatory authority for use in this emergency. Some countries in the middle east such as Bahrain and the United Arab Emirates have approved the usage of BBIBP-CorV and Russia has allowed Sputnik V to be used widely while ignoring the consequences and safety protocols \cite{Cyranoski2020}. Despite the availability of data, the answers to unknown variables remain at large. For instance, the duration of immunity, viral transmission, the safety protocols for a vulnerable group such as pregnant women and elders, and the risks associated with the vaccine’s adverse reactions \cite{TheLancetMicrobe2020}. The availability of the data regarding these questions is limited to none. Still, the question regarding the logistics, financial, and social implications are not touched upon. According to Anthony Fauci (Head of the National Institute of Allergy and Infectious Diseases), the only prominent and long-term solution to this pandemic is to follow the public health guidelines which include frequent hand wash, indoor activities, physical and social distancing, wearing masks, and enhancing immunity through natural means \cite{Schmidt2020}. It has also been concluded by a recent study \cite{Wise2020,TheLancetMicrobe2020} that even though the vaccines will play a vital role in controlling the pandemic but the immediate effects will not be observed at all due to the challenges regarding global distribution and lack of data availability. Furthermore, Jonathan Samet (MD dean of Colorado school of public health) in a recent interview suggested that even though an individual gets vaccine, that individual will not be allowed to physically or socially interact as well as the condition of wearing masks still holds, therefore, hygiene, distancing, and masking needs to be prioritized \cite{Mccoy2020}. As per the Colorado department of public health and environment, it is still possible for an individual to get mild infection even after vaccination and that they might still be the transmitters of virus \cite{col}.  
It is evident from the facts provided that the long term setting for dealing with this pandemic is to slow down the transmission by changing our lifestyle and complying with the public health guidelines. Unfortunately, it has been observed that the guidelines are not being followed by the general population. Furthermore, a global survey conducted from 19 countries recently concluded that 71.5$\%$ of people have shown acceptance for the COVID-19 vaccine provided that their employer or the government recommends it \cite{Lazarus2020}. The study also states that the acceptance ratio is higher, i.e. 90$\%$, where the people have trust in their respective governments such as (China, South Korea, and Singapore) but the acceptance drops to 55$\%$for the countries like Russia. One more thing to consider is that the people who showed acceptance rely on the recommendations of the employer or the country which has not been taken into account for the acceptance model. Considering the constraints of logistics, mutations, and vaccine acceptance, personal and public monitoring systems are in dire need which not only helps to provide recommendations regarding the safety guidelines and hygiene but also helps to improve the lifestyle and immune system at the same time. One way to accomplish the aforementioned task is to perform continuous monitoring while collecting huge amounts of data. With the advent of microelectromechanical systems (MEMS) the sensors have evolved to be small in size as well as effective. The miniaturization of the sensors has allowed the production of wearable sensors that can collect large amounts of data as well as help in improving an individual’s health. Furthermore, combining the use of wearable and miniaturized sensors with ICT technologies such as 5G/6G communication, Big Data, Artificial Intelligence, and the Internet of Everything will be able to improve healthcare services, personal hygiene, immunity boost, mental healthcare, and contact tracing problems, accordingly. It has been proven by many studies that wearable devices and mobile sensors, are widely accepted by the audience varying from children to the elders \cite{Dai2020,Li2019,Wang2020,Dutot2019}. The integrated services will help to alleviate the COVID-19 and future pandemic related issues on a personal level which will eventually lead to facilitate the financial issues of a country on a macro level. 
This article focuses on the use of the Internet of Everything (Wearable sensor analytics) to deal with physical healthcare monitoring, personal hygiene and immunity-boosting, Mental healthcare, and Contact tracing which are considered to be good practice adaptations to slow down the transmission of infection. We provide possible implementation strategies, the challenges, and potential solutions for each of the aforementioned issues. To the best of our knowledge, a framework in compliance with the Internet of Everything guidelines for long term setting to deal with current and future pandemics has not been provided. Particularly, the contributions of this study are given below:
	\begin{itemize}
    \item Realistic process flows or frameworks for societal issues based on the wearable sensor characteristics.
    \item Analyzing the limitations and challenges for each of the societal issues from the perspective of wearable sensor analytics. 
    \item  We propose the Covid-19 Resistance Framework using the Internet of Everything (CORFIE) to help align the good practices in this \textit{new normal}. 
\end{itemize}
The rest of the paper is structured as follows: We present preliminaries for Internet of Everything based architectures and the layers involved in Section 2. We propose hypothetical frameworks for physical healthcare monitoring, personal hygiene and immunity-boosting, mental healthcare, and contact tracing in Section 3, 4, 5, and 6, respectively. Section 7 presents the details regarding the proposed CORFIE. Section 8 highlights some potential challenges and issues when adapting CORFIE architecture and Section 9 concludes this study. 

\section{Preliminary Understanding}
The following subsections will extensively use some terminologies which are specific to Internet of Everything. The IoE is derived from Internet of Things (IoT) which combined the network and the things. However, IoE extends the relationship and network and things with people, data, and processes. Some studies referred IoE as Internet of Things, Services, and People (IoTSP) [18]. The core idea and architecture used behind IoE and IoTSP is the same. As we present realistic frameworks for the prevention of COVID-19 and future pandemics, we define some terminologies related to IoE, accordingly. It should also be noticed that the presented frameworks and the layers are in line with the IoTSP architecture.

\begin{itemize}
    \item \textbf{Sensor Layer:} This layer either acts as a standalone or a collection of memory-constrained, small, and battery operated sensing devices. The scope of IoE expands the scope of this layer and considers it as a collection of devices that is connected via Internet and is capable of sensing data.
    \item \textbf{Access/Communication Layer:} As the name suggests, this layer comprises of set of communication protocols that can relay the data from sensor devices to the middleware or directly to the server. The communication protocols include but are not limited to Sub-GHz proprietary, long range (LoRa) WAN, Zigbee, Bluetooth, radio frequency (RF), long term evolution (LTE), LTE-advanced (LTE-A), and Wireless Fidelity (WiFi) Direct.
    \item \textbf{Middleware:} This layer is mostly considered as a software but in this study we consider the smartphone as the middleware which acts as an intermediary for collecting the data from sensor layer and sending it to the server layer. Furthermore, the middleware can be used for data management, API management, authentication, messaging, and application services, respectively. In the present architectures, middleware is also responsible for fetching the decisions related to specific services from server layer and send it to the notification center. 
    \item\textbf{Server Layer:} Although the server layer varies with respect to the evolution of IoT studies. However, the common aspect of this layer is to receive the data from middleware, process, analyze, and store it to provide specific services. In this study, server layer comprises of two components, i.e. context (bag of contexts) and services (pool of services). The context can either be recognized automatically or derived from statistical inferences (implicit service). The context selection needs to be performed in order to select explicit services (pool of services), accordingly. The service layer is also equipped with decision making and data analytic techniques.
    \item \textbf{Notification Center:} Once the middleware acquires decision based on the explicit services, it is pass to the notification center in form of alerts and recommendations. The notification center also allows the user to visualize the measurements obtained from sensing devices in real-time. 
\end{itemize}	

The flow of the architectural layers presented in the subsequent sections is given below:

\begin{enumerate}[label=\alph*.]
\item The data is acquired from the sensor layer and sent to middleware via access/communication layer.
\item The middleware stores the data temporarily acquired from the sensor layer via access/communication layer.
\item The temporarily stored data is accessed by the context (a component in the service layer) to determine the context. (The determination of context is an implicit service which will be carried out without the user preference or intervention).
\item Based on the selected context, an individual service or pool of services will be activated and perform the desired task.
\item Middleware acquires the decision obtained from the activation of desired service(s).
\item Middleware sends the result of the activated service to the notification center.
\end{enumerate}

\section{Physical Healthcare Monitoring}
Monitoring of physical human health is of utmost importance during this pandemic especially if an individual is self-isolated in case of getting infected or due to voluntary isolation after getting vaccinated. It is also a way to offload the bottleneck from the hospitals or healthcare provision institutes that are overwhelmed with the inflow of patients even with mild symptoms. We present a summary of the embedded sensors with smart devices (not limited to) in Table 1. The increase in embedded sensors leads to an increase in price as well. However, some good and not so expensive smartwatches can be used for basic physical healthcare monitoring. Furthermore, some of the digital instruments which monitor the physical health of a patient in an individual manner can also be used for monitoring purpose. There are many aspects of physical healthcare that can be monitored using a smartwatch in connection with smartphones such as, physical activity, skin temperature, oxygen saturation, and anomaly detection. Physical activity is an important aspect that not only can be used to check the patient’s activities but also helps in boosting immunity. We will shed light on the immunity-boosting using physical health in the subsequent section. In this section, we focus on the monitoring of physical activities while in isolation and its compliance with the routine suggested by the caregiver/doctor. \\
The inertial measurement units such as accelerometer, gyroscope, magnetometer, and orientation sensors embedded in smartphones, as well as smartwatches, has been used extensively for physical activity recognition over the years \cite{Khowaja2018,Khowaja2017,Khowaja2016,Khowaja2020a}. The problem with the adaptation of activity monitoring approaches is the lack of personalized data and the variation in wearable sensor devices in terms of sampling rate, recording units, and so forth. Some studies solve the problem to an extent by either using a semi-population calibration approach so that personalized activity recognition could be carried out with a minimum level of annotated data, and transfer learning approaches that take into account the domain adaptation of varying sensor characteristics. Recently, a study \cite{Khowaja2020a} proposed the use of behavior for personalized activity recognition using a semi-population calibration approach which not only considers the diverse nature of human behavior when performing certain activities but also the varying class labels. Physical activity recognition will allow the monitoring of patients in an automated way without physical interaction with the caregiver/doctor. \\
Skin temperature and oxygen saturation monitoring are of utmost importance with regards to the COVID-19 pandemic. As COVID-19 (SARS-nCOV-19) is a respiratory syndrome it causes high fever as well as breathing issues to the infected individual. The fever and respiration issues are the most common symptoms of COVID-19 virus contraction listed by the world health organization. Considering that the skin temperature and the oxygen saturation (SpO2) can be monitored continuously through the smartwatches is a relief to both the patients as well as doctors. In the isolation phase, the spike in skin temperature or decline in oxygen levels might recommend the patient to take necessary action or notify the concerned doctor for immediate response in time. Furthermore, the sensor modalities can be used to log the temperature and oxygen responses in case of voluntary isolation after the vaccination. \\
The field of anomaly detection using wearable sensors has been in discussion for quite some time. In many countries, elders prefer to live alone which make them vulnerable to some anomalous activities such as fall or slip and also allow them to skip their medications due to memory loss. The anomaly detection in the isolation phase can be used to monitor the same anomalous activities which might alert the authorities in case of sudden fall that require immediate attention or recommend an individual about their medication which needs to be taken at a specified time. There are many studies which propose the use of smartwatch and smartphones for detecting falls accurately \cite{Khowaja2018,Khowaja2016}. Furthermore, the isolation center or the place where individuals get isolated can be equipped with an object, infrared, or location sensors, to detect anomalous behaviors as well such as sleeping, eating, and physical behaviors that may help the caregivers to understand more about the progression of the virus, accordingly. \\  
We propose a hypothetical framework, physical healthcare monitoring module (PHM) as shown in figure 1. The framework is compliant with the sensor devices shown in table 1 except the beacons which could be used as location sensors. The smartphone acts as a middleware that will be responsible for sending the data acquired from sensors to the service layer for specific service provision. The smartphone is also responsible for fetching the decision obtained from the service layer and display it to the mobile application interface or send the decision to the doctor/caregiver for necessary action. The steps indicated by the small letters in figure 1 are compliant with the steps mentioned in Section 2. The PHM is one of the modules which could help in monitoring a patients’ health while in the isolation phase which is an essential phase if one gets infected or vaccinated alike. Furthermore, the module can transform even an individual room of a distant house into an isolation ward in terms of healthcare service provision.
\begin{table*}[]
\caption{List of some smartwatches embedded with wearable sensors}
\centering
\label{tab:my-table1}
\renewcommand{\arraystretch}{1.2}
\begin{threeparttable}
\begin{tabular}{|l|l|l|l|l|l|l|l|l|}
\hline
Company \& Device Name        & ST & BR & GSR & IMU & HR & SpO2 & Other               & Price* \\ 
\hline
Honor Magic Watch 2   & \xmark  & \xmark    & \xmark & \checkmark     & \checkmark     & \xmark   & SLQ   & \$171 \\ 
\hline
Samsung Galaxy Watch Active 2 & \xmark  & \xmark    & \xmark & \checkmark     & \checkmark     & \xmark   & SLQ   & \$330 \\ 
\hline
Garmin Fenix 6                & \xmark  & \xmark    & \xmark & \checkmark     & \checkmark     & \xmark   & SLQ   & \$550 \\ 
\hline
Huawei GT 2              & \xmark  & \xmark    & \xmark & \checkmark     & \checkmark     & \xmark   & -   & \$170 \\ 
\hline
Apple Watch Series 1         & \xmark  & \xmark    & \xmark & \checkmark     & \checkmark     & \xmark   & SLQ   & \$280 \\ 
\hline
Fitbit Sense                 & \cmark     & \cmark   & \cmark     & \cmark     & \cmark    & \cmark     & AFib, SLQ           & \$280  \\ \hline
Samsung Galaxy Gear S3        & \xmark    & \xmark    & \xmark    & \cmark    & \cmark   & \cmark      & SLQ                 & \$310  \\ \hline
Biobeat Smartwatch            & \cmark    & \cmark    & \cmark    & \cmark     & \cmark    & \cmark      & BP, CI, MAP, SL, SV & \$2800 \\ \hline
Biostrap EVO                  & \xmark    & \cmark   & \xmark    & \cmark     & \cmark    & \cmark      & SLQ                 & \$250  \\ \hline
Empatica Embrace              & \cmark   & \xmark    & \cmark     & \cmark     & \xmark    & \xmark      & SLQ                 & \$250  \\ \hline
Empatica E4                   & \cmark    & \xmark    & \cmark    & \cmark     & \cmark    & \cmark      & SLQ                 & \$1640 \\ \hline
Xiaomi Mi Band 5              & \xmark    & \xmark    & \xmark     & \cmark     & \cmark    & \cmark     & SLQ                 & \$50   \\ \hline
Realme Watch S                & \xmark    & \xmark   & \xmark    & \cmark     & \cmark    & \cmark      & SLQ                 & \$94   \\ \hline
\end{tabular}
 \begin{tablenotes}
   \small
    \item *Price may vary, ST $\rightarrow$ Skin temperature, BR $\rightarrow$ Breathing rate, GSR $\rightarrow$ Galvanic skin response, IMU $\rightarrow$ Inertial measurement unit, HR $\rightarrow$ Heart rate, SpO2 $\rightarrow$ Oxygen saturation, SLQ $\rightarrow$ Sleep quality, AFib $\rightarrow$ Atrial fibrillation, BP $\rightarrow$ Blood pressure, CI $\rightarrow$ Cardiac index, MAP $\rightarrow$ Mean arterial pressure, SL $\rightarrow$ Sweat level, SV $\rightarrow$ Stroke volume. 
    \end{tablenotes}
 \end{threeparttable}
\end{table*}

\begin{figure*}[ht!] 
\centering
\includegraphics[width=5.2in]{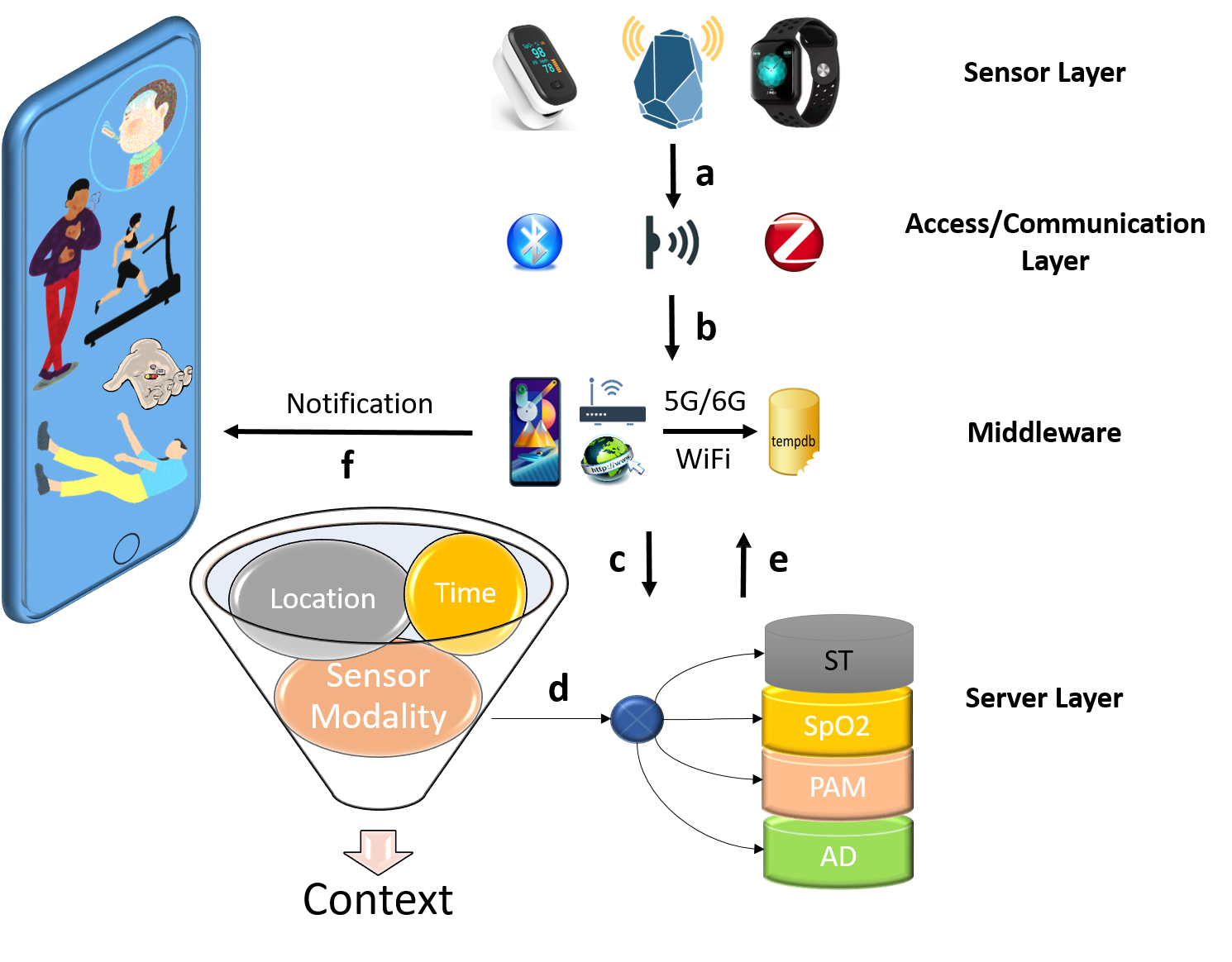}
\caption{The physical healthcare monitoring module for patient monitoring during the isolation phase. ST $\rightarrow$ Skin temperature, SpO2 $\rightarrow$ Oxygen saturation, PAM $\rightarrow$ hysical activity monitoring, AD $\rightarrow$ Anomaly detection.}
\label{fig:architecture}
\vspace{-5.5mm}
\end{figure*}

\section{Personal hygiene and Immunity boosting}
According to the world health organization's public safety guidelines with regards to the COVID-19 pandemic, personal hygiene is of utmost importance in order to slow down the spread of viral infection. Personal hygiene includes washing hands with water and soap for more than 20 seconds, use of sanitizer in case the former option is not available, repeating the steps if cough or sneeze is detected, accordingly. Furthermore, it’s not just sneezing or coughing that releases the droplets, even talking to someone or having a conversation can release the droplets as well. Therefore, having a recommender system that reminds an individual for taking care of personal hygiene can play a vital role in reducing viral transmission. 
When it comes to wearable sensors, personal hygiene can be monitored and recommendations can be suggested using a camera from smartphones, a microphone from smartphones as well as smartwatches and the inertial measurement units. The IMUs in smartwatches will help in detecting the washing hands and brushing teeth actions, and the recommendations might be provided at specific timelines to repeat the respective actions. The smartphone camera could be used to scan the barcode of the sanitizer in case the option of washing hands is not feasible. The barcode will provide the information to the users about the alcohol level and recommend its usage as per the WHO guidelines. The crucial part of such a recommender system related to personal hygiene will be based on microphones available in smartphones and smartwatches. The acquisition of sound waves can distinguish between the cough, sneeze, and normal sounds. In case of cough and sneeze sound detection, the user can be provided with the recommendation of washing hands again or using sanitizer immediately. Furthermore, as per WHO recommendation, the distance of the droplets from cough or sneeze depends on several factors such as a person having a full set of teeth, blocked nose, larger viral load, and louder voice. These factors determine an individual as a super spreader as their sneeze can travel up to 60$\%$ further and can produce 4 times more droplets \cite{Chaudhuri2020,Fontes2020}. If such information can be gathered through a profiling system, the recommendation based on user-profiles and the microphone sensor can be made regarding the physical distance that needs to be maintained. Another aspect of maintaining personal hygiene is proper ventilation and maintaining good air quality at the place of isolation. Many mobile sensors can be directly paired with smartphones to provide information regarding air quality so that a feasible recommendation can be provided. 
A crucial aspect of being probably safe during this pandemic is to boost the immune system through healthy foods and diet. Students have proved that the use of water in large quantities, the use of zinc and magnesium, foods rich in vitamins C, D, and E, herbs, and the use of some specific ingredients can improve an individual’s immune system. However, it is not always possible for everyone to consume such enlisted food items for immunity boosting due to allergies, health constraints, and so forth. Most of the food items in the marts have a barcode or QR code that provides the information related to the ingredients used in the product. The camera in a smartphone can easily capture the bar or QR code to scan the items used in the product and provide a recommendation of its usage in compliance with the health standards and the individual’s profile so that the respective allergies and health issues can be kept in check. Several smartphone apps provide such information while scanning the aforementioned codes such as myfitnesspal \footnote{https://www.myfitnesspal.com/ }and more. A question can be made that homemade foods do not come with any bar codes or QR codes. In that case, the natural language processing characteristics can be used to search the ingredients in a certain food item online in an automated way as well as providing a voice command to search for an immunity-boosting recipe, accordingly. The use of microphones and the keyboard input either from a smartphone or smartwatch can be leveraged for the said purpose. Google’s speech recognition and search packages in python help in performing such tasks \footnote{https://github.com/sander-ali/News-Scraper}. As mentioned earlier, physical activities can also help in boosting immunity. According to studies \cite{Khoramipour2020,Damiot2020,Vorvick2020}, physical exercise can help flush bacteria out of airways and lungs which reduces the chances of catching flu, cold, or other bacterial infections. Physical exercises can also help in changing white blood cells and antibodies which are an essential aspect in dealing with viral diseases. Similar to the food items, all physical exercises are not meant to be performed for individuals due to their body types, health constraints, and other issues. Based on an individual’s profile, the system can recommend the type of exercises that are considered to be safe.  
The hypothetical framework for personal hygiene and immunity-boosting (PHIB) module is shown in figure 2. The framework only opts for the wearable and mobile sensors that could be connected via smartphone for data acquisition. The framework is also compliant with the Internet of Things (IoT) layers for real-life applicability. The data from multiple sensors is acquired and stored in a temporary database in the middleware. The camera can help in reading bar and QR codes, respectively. The microphone sensor can be used to detect cough or sneeze sounds as well as recording users’ queries for relevant exercises and food ingredients. The GPS sensor in the smartphone helps recognize the context as the person is indoor or outdoor. The indoor location context will activate a certain set of services such as recommendations for indoor exercises, cooking, hand wash by detecting coughing and sneezing sound at home, air quality index, and more, while the outdoor location triggers physical distance alert, foods available in restaurants, sanitization alert, and outdoor exercises. The data is then pushed to the server for further processing, detection, recognition, and recommendation, accordingly. The user gets notified on their mobile screens for the desired service.

\begin{figure*}[ht!] 
\centering
\includegraphics[width=5.2in]{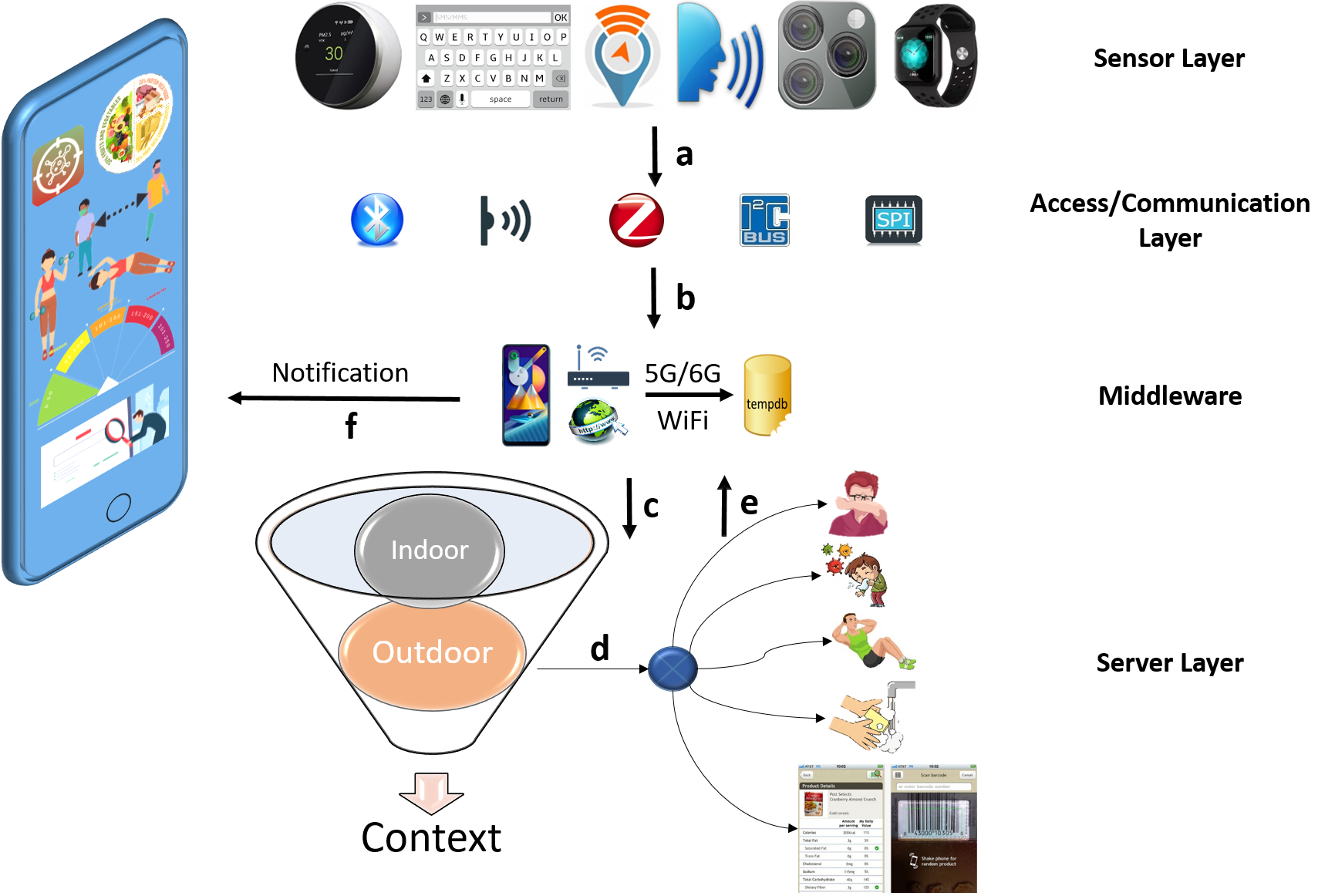}
\caption{Personal hygiene and immunity-boosting module for indoor and outdoor locations.}
\label{fig:architecture}
\vspace{-5.5mm}
\end{figure*}

\section{Mental Healthcare}	
A surge in stress, anxiety, and depression has been observed as an after effect of the COVID-19 pandemic. Although the pandemic is not over yet the stress and anxiety continue to play a causal role in mental health conditions. Researchers have even related the emotional and behavioral response of individuals in this pandemic to the terror management theory where the fear of death plays an important part in making daily life decisions \cite{Fofana2020,Menzies2020,Madigan2020}. Furthermore, many studies have concluded that self-isolation may lead to prolonged stress which might result in anxiety, distress, or depression, depending on the length of the isolation period. Studies have also suggested that the fear and anxiety in the times of COVID-19 have increased the crisis on public healthcare services. Recently center for disease control and prevention (CDC) had also emphasized managing stress and anxiety levels as stress during the infectious period can result in abnormal eating and sleep patterns, an increase in chronic health issues, reduced immunity, and panic attacks \cite{NationalCenterforImmunizationandRespiratoryDiseasesNCIRD2020}. According to CDC, elders, front-line health workers, socially isolated, people who lost their loved ones, people who lost jobs, and people facing financial crisis might react strongly to stress crisis during the pandemic. Another recent study \cite{Sher2020} suggested that as the stress gets prolonged, it could result in suicidal behavior if not intervened with a timely and relevant response. Reports from CDC, John Hopkins University, and Boston university of public health have concluded that anxiety and depression have been tripled and quadrupled this year in comparison to the preceding years \cite{Ettman2020,McGinty2020}. 
According to CDC, coping with stress will not only be helpful for an individual’s health but it would affect the overall community in terms of extended empathy, voluntary support, increased social connection, and less strain on public health services. However, in achieving community-wide benefit the emotional distress needs to be diagnosed in time. Mental health disorders such as emotional distress and stress can be measured with wearable sensors including electrocardiograph (ECG), GSR, and electroencephalogram (EEG) \cite{Khowaja2018,Setiawan2018,Khowaja2020}. Some of the smartwatches listed in Table 1 comprise of such sensors that could help in recognizing the emotions of an individual, accurately. However, mobile sensors for such modalities are also available commercially which could be connected with smartphones for continuous data collection. Timely stress recognition may notify doctors/caregivers, family members, friends, loved ones, and volunteers to help relieve the stress of an individual. Moreover, timely detected stress can also be handled with home automation as COVID-19 is not the only cause for inducing anxiety. Some studies have also considered the usage of smartphones and posts on social media to determine stress levels. The use of natural language processing can be leveraged for determining stress as well while considering text or voice conversation with the chatbot as an input.	

A hypothetical framework for the mental healthcare (MHC) module is presented in Figure 3. The middleware can acquire the data from the wearable and mobile sensors on a continuous basis and store the data temporarily in the buffer for some seconds before sending it to the server. The server will be able to detect anomalous emotions such as anxiety, depression, distress (prolonged stress), and suicidal tendencies. The context of isolation and non-isolation is selected based on the studies \cite{Fofana2020}, \cite{Madigan2020}, \cite{Ettman2020}, \cite{McGinty2020} which concluded that the stress related to self-isolation, infection in COVID-19 pandemic has a different characteristic than that of financial, job, and occupational stress. The context of pre-and post-COVID infection can also be taken into account for the MHC module. Based on the detected anomalous emotion multiple actions/recommendations can be provided such as variations in lighting conditions, calling to family or friends, having a conversation with an emotional chatbot, and so forth. Furthermore, this system can also be employed for stress in people driving cars or with long commute times for suggesting them an alternative fast route. 

\begin{figure*}[ht!] 
\centering
\includegraphics[width=5.2in]{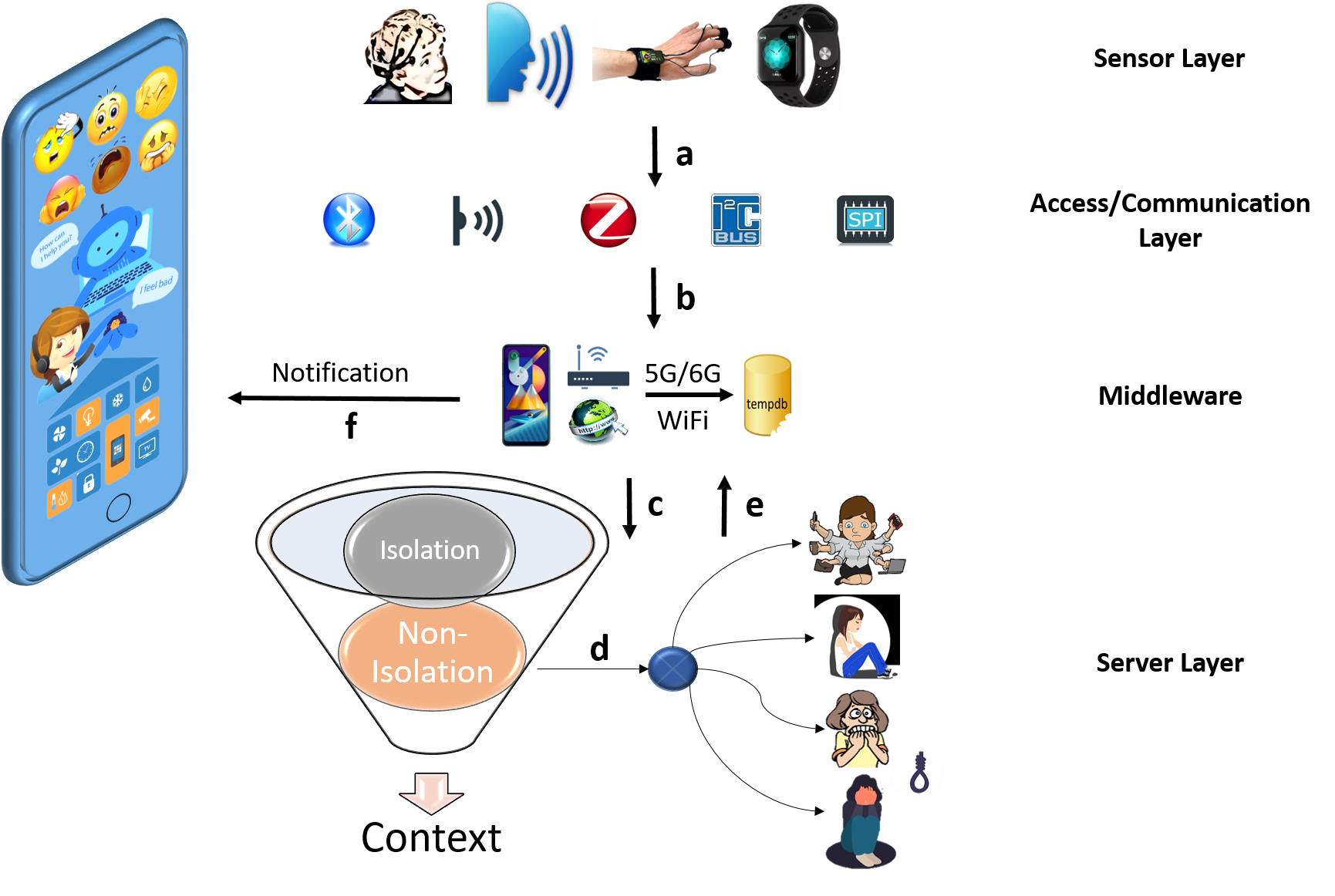}
\caption{Mental Healthcare module for isolation and non-isolation phases.}
\label{fig:architecture}
\vspace{-5.5mm}
\end{figure*}

\section{Contact Tracing}
The importance of contact tracing was highlighted when “patient 31” in South Korea infected hundreds of patients over the span of days. With the outbreak of COVID-19 manual contact, tracing procedures were performed by the health departments of different countries. The process is time extensive, prone to errors, and is not scalable, thus, appears to be ineffective so far. Since then, government officials and policymakers have moved towards a digital solution based on intelligent decision support systems. Recent studies \cite{Chowdhury2020,Ahmed2020,Kretzschmar2020} have urged the use of smartphones and mobile sensors for contact tracing in order to limit the transmission of the virus, hence resulting in smart lockdowns. Although solutions have been provided for proximity tracing and alerts, the roll-out plan and tracing of contacts have been limited due to mainly two reasons. The first is the tracing only using GPS sensors using smartphones and the second is the lack of behavioral consideration. The first-ever contact tracing protocol for dealing with the COVID-19 pandemic was developed by Singapore’s government which used Bluetooth to perform the respective tracing. The protocol was based on OpenTrace \footnote{https://bluetrace.io/}, an open-source implementation of Android and iOS apps along with baseline calibration data and cloud server backend. The protocol can be called by its alternative terms such as BlueTrace and TraceTogether. The Australian government in April 2020 launched an app that traces the contact manually that complies with the BlueTrace protocol and was termed as COVIDSafe \footnote{https://www.health.gov.au/resources/apps-and-tools/covidsafe-app}. The Chinese app (Chinese health code system), South Korean app (Corona 100m), and the United Kingdom’s (NHS COVID-19) app has been proposed with similar characteristics. There have been other studies that proposed the contact tracing algorithms but face the same fate of being ineffective when it comes to scalability. A recent study \cite{Khowaja2020a} highlighted the importance of behavior analysis when it comes to activities of daily life. The use of behavior analysis can be incorporated with the activity recognition using wearable sensors to derive context rather than just tracing the persons gathered at a common place. A person who works as a waiter in a restaurant performs an activity and exhibits behavior differently than the one who visited for having lunch or dinner. Both of them have different levels of exposure while coming in a contact with an infected individual. Furthermore, the use of natural language processing can be garnered in order to get additional information about the places where an individual has traveled. Furthermore, the ones who visited the probable infected places can be notified in the same time span. The activity combined with the given context can reveal whether the person has just passed by a certain restaurant in a vehicle or by walk which again provides a different perspective of infection spread via an individual. 

With the emergence of new COVID-19 variants and mutations, the importance of contact tracing has been raised manifold. It is a necessity to keep the track of individuals that have traveled back and forth from a certain location where the variants are prominently discovered. Moreover, such a kind of contact tracing will help to identification of vaccine variants that need to be injected with respect to the location of the mutation. The GPS sensor along with inertial measurement units, speech analysis, and social media content, the automation of contact tracing could be improved with real-life applicability.

\begin{figure*}[ht!] 
\centering
\includegraphics[width=5.2in]{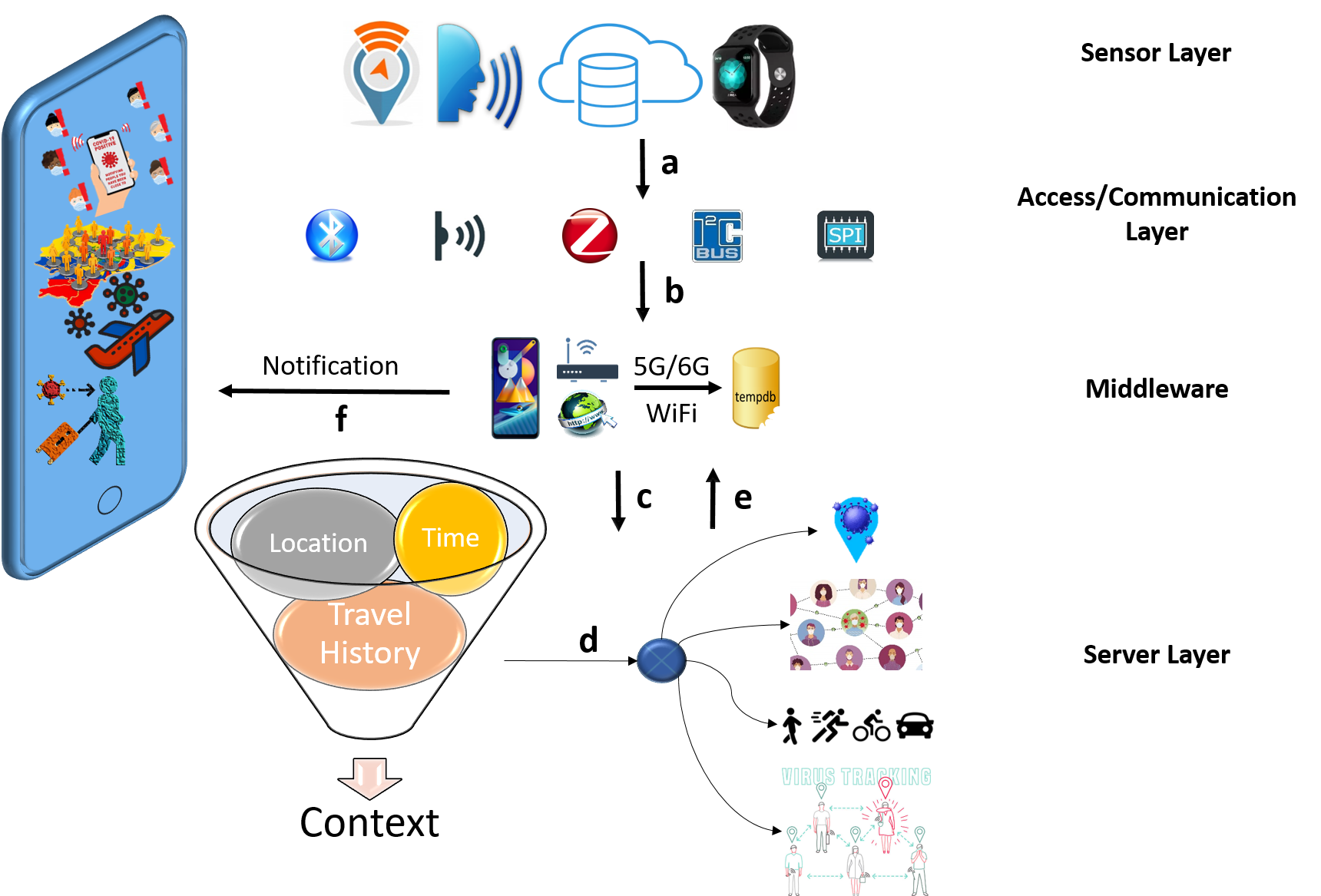}
\caption{Contact Tracing Module based on the location, time, and travel history.}
\label{fig:architecture}
\end{figure*}

We present a hypothetical framework for the contact tracing (CTC) module in figure 4. The CTC block considers the data from GPS sensors, inertial measurement units, voice input, and travel history of individuals using the system. The data is temporarily stored in the middleware and is pushed to the server for necessary detection based on the location, time, and travel history contexts. The location and time context have been used to integrate behavioral characteristics when recognizing high-level activities, therefore, the recognition of activities at specific locations can provide insights of a person either infected or not. With the unified location history of the individual using the system, contact tracing can be made easy based on the intersection of locations. Furthermore, in the case of a person traveling from a specific area where a new mutation or variant of COVID-19 has been found, an alert can be generated to users in the travelers' proximity. The system can also highlight locations where the COVID-19 infection risk is high so that the users may avoid the route or place altogether. An additional feature of distance measuring can be added between the smartwatches of different people to recognize handshake activity which could help trace the users in case of either one being infected in prior. 

\section{COVID-19 resistance framework using IoE}
Based on the four hypothetical modules, we propose a general hypothetical COvid-19 Resistance Framework using the Internet of Everything (CORFIE). It can be noticed from the modules that more-or-less the sensor, access/communication, and middleware layer are similar. However, the context and server layer vary with respect to the targeted module. Furthermore, a general dashboard in an app could be constructed for either automatic or manual activation of a particular service. The proposed COFIE framework is shown in figure 5. We briefly define each of the layers in the CORFIE framework.

\begin{figure*}[ht!] 
\centering
\includegraphics[width=5.2in]{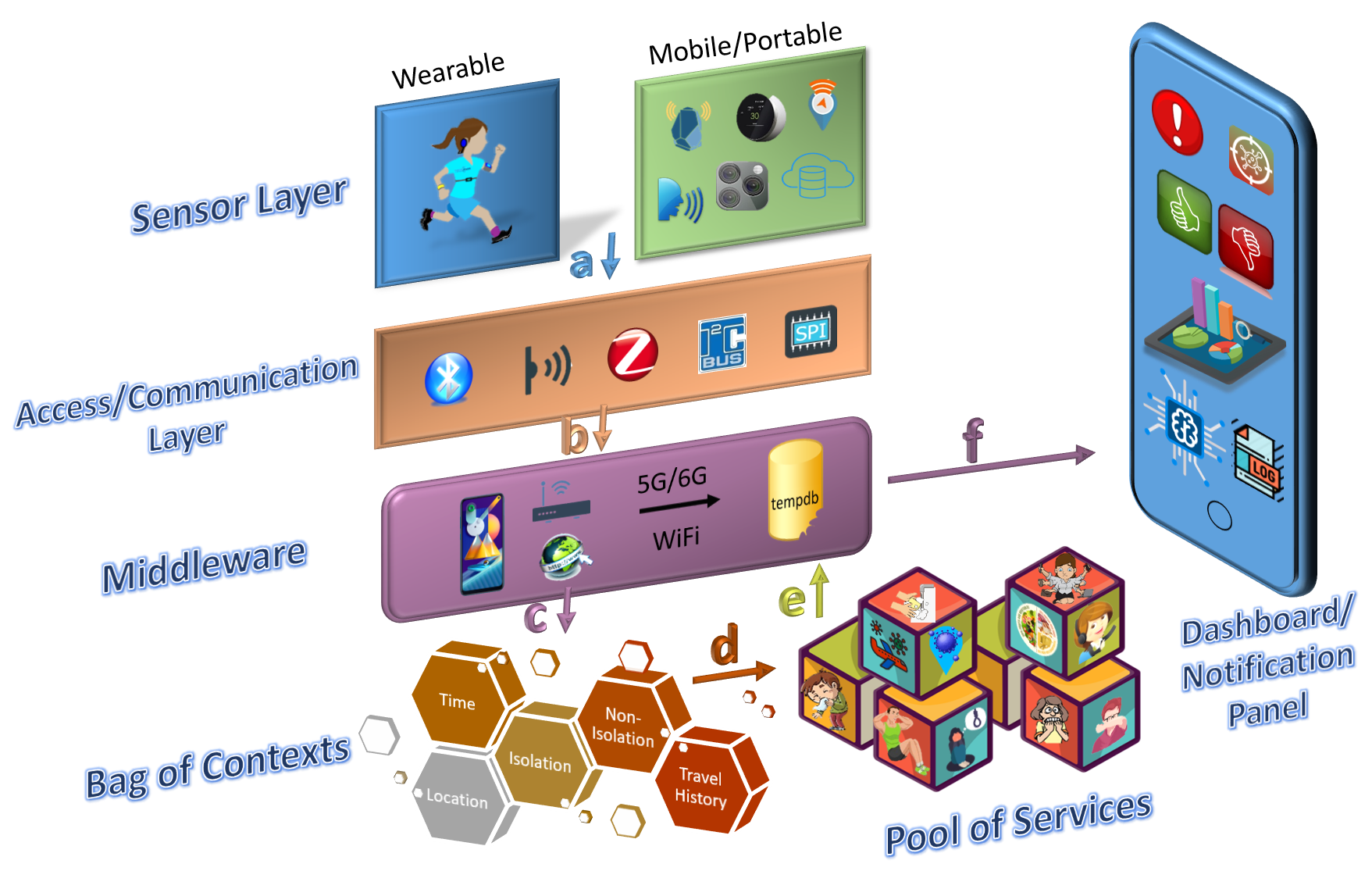}
\caption{Proposed Covid-19 Resistance Framework using the Internet of Everything (CORFIE)}
\label{fig:architecture}
\end{figure*}

The sensor layer in the CORFIE framework is responsible for data acquisition from various wearable and mobile sensors. The only constraint we put on the selection of the sensors is their ubiquity and pervasiveness. Although a single sensor device may house multiple sensors, for instance, an inertial measurement unit might comprise of accelerometer, gyroscope, magnetometer, and other sensors as well, but the sensor layer of CORFIE framework considers each of the sensor measurement separately despite the device used or embedded in. The reason for considering individual sensor measurements is due to the diversified characteristics of embedded sensors and associated applications. For example, the breathing sensor can be used for anomaly detection in an isolation phase whereas the same sensor is used for physical activity or exercise recognition in the non-isolation phase. In this regard, the CORFIE framework assumes that there are $\tau$ sensor devices, i.e. \emph{DEV}= \{dev$_t$ $|$ t=1,....,$\tau$\} and $\rho$ sensor types in each dev$_t$, i.e. SD= \{sd$_p$ $|$ p=1,....,$\rho$\}. Using the sensor device and its types we can define an individual sensor measurement as shown in equation 1.
	\begin{equation} 
IMS_\epsilon = \Bigg \langle IMS - id_\epsilon, dev_t,_\epsilon, sd_p,_\epsilon \bigg \rangle
	\end{equation}

where IMS$_\epsilon$ refers to the individual measurement of sensors, IMS - id$_\epsilon$ is the unique id for each individual measurement from sensors, and dev$_t,_\epsilon$, sd$_p,_\epsilon$ are the sensor device and sensor type, respectively.
The acquired IMS will then be transferred to the middleware via the access/communication layer. This layer acts as a source of communication between the sensor layer and middleware by using abstract protocols such as serial peripheral interface (SPI), inter-integrated circuits (I2C), Zigbee, infrared, and Bluetooth. The CORFIE is a general framework, therefore, the protocols in the access/communication layer are not limited to the aforementioned ones and can be extended to other protocols such as RFID, depending on the given context and service. 
As the CORFIE framework is mainly designed considering the wearable and mobile sensors, therefore, the smartphone is considered to be the middleware that acts as an intermediary device for all the layers, respectively. The data acquired from the sensor devices via the access/communication layer is stored temporarily in the memory which then is pushed to the server via a gateway or 5G/6G services. Once the data is pushed, new data acquired from the sensors will be overwritten, accordingly. It has been proved in the existing studies that the storage of data in memory temporarily is a light-weight operation \cite{Khowaja2018}. The pre-processing of IMS$_\epsilon$ for opting context and triggering desired service will be performed on the server layer. The decision from the desired service is then collected from the middleware while using REST API endpoints to call a specific function. The decision is then either shown to the application dashboard or sent to the doctor/caregiver, respectively. 
The bag of contexts is of vital importance and is considered to be an implicit service in the CORFIE framework. The consideration of context from IMS$_\epsilon$ can be performed either using knowledge-based or data-driven approaches. Some studies also considered the use of hybrid methods which leverages the characteristics of both the knowledge- and data-driven approaches for recognizing the context in an automated manner. The knowledge-driven approaches use resource description framework (RDF), web ontology language (OWL), and simple protocol and RDF query language (SPARQL), whereas the data-driven approaches use machine learning techniques to perform the desired task. The bag of contexts in the CORFIE framework is represented as BoC$_m$ = \big\{c$_m$ $|$ m=1,...,M\big\} where M represents the number of available contexts. 
Once the individual measurement from sensors is obtained the data could be sent to the pool of services for activation based on the selected context. The pool of services might include a web-server for a decision support system, a self-designed app in a smartphone, or a third-party app such as Samsung Health, myfitnesspal, and so forth. Furthermore, the pool of services will implicitly store the data for summarizing the data, provision of logs, and alerts for the use of medical assistance. The pool of services in the CORFIE framework is defined in equation 2

	\begin{equation} \label{eqn2}
	PoS_n=\big\{ pos-id_n, IMS_\epsilon,_n,BoC_m,_n\big\}
	\end{equation}

The PoS$_n$, pos-id$_n$ refer to the selection of a specific service and its unique id, respectively. It should be noticed that similar context can be associated with multiple services, for instance, location and time contexts are considered for the services listed in PHMM and CTC modules. Therefore, multiple services can be triggered with respect to the selected contexts. A search algorithm for initializing a particular service is given in table 2. The algorithm looks into all the services available (Line 2), lists the services with similar context requirements (Line 4), and for all similar services (Line 5) checks the availability of the sensor measurements (Line 6). In the case of available measurements, the CORFIE will activate the desired service, respectively.  

\begin{table}[]
\caption{Search algorithm for service selection in CORFIE framework}
\centering
\label{tab:my-table}
\begin{tabular}{|l|l|}
\hline
  & Search Algorithm                    \\ \hline
1 & Initializing Pool of Services       \\ \hline
2 & PoS = \big\{pos$_n$ $|$ n=1,...,N \big\} \\ \hline
3 & For pos$_n$ in  PoS                             \\ \hline
4 & CSV = \big\{pos$_z$ $|$ z=1,...,Z \big\}, where Z $\leq$ N and CSV $\subseteq$ PoS                        \\ \hline
5 & For pos$_z$ in CSV                              \\ \hline
6 & Check IMS$_\epsilon$,$_n$                               \\ \hline
7 & \textbf{IF} sensor measurements available \\ \hline
8 & Activate the service             \\ \hline
\end{tabular}
\begin{tablenotes}
\item IMS$_1$\ = \big \langle 1, 1, Acc \big \rangle, IMS$_2$ = \big \langle 2, 1, Gyr \big \rangle, IMS$_3$ = \big \langle 3, 2, BR \big \rangle,\\ IMS$_4$ = \big \langle 4, 3, HR \big \rangle, IMS$_5$ = \big \langle 5, 4, Maps \big \rangle \\
BoC$_1$ = \big \langle Time and Location \big \rangle \\
PoS$_1$ = \big\{1, \big \langle IMS$_1$, IMS$_2$, IMS$_3$, IMS$_4$ \big \rangle ,\big \langle Time and Location \big \rangle \big\} \\
PoS$_2$ = \big\{2, \big \langle IMS$_1$, IMS$_2$, IMS$_5$ \big \rangle ,\big \langle Time and Location \big \rangle \big\}
    \end{tablenotes}
\end{table}
In the above example, five individual sensor types from four sensor devices have been acquired. Based on the Time and Location context two services, i.e. physical activity monitoring (PoS$_1$) and COVID-19 tracking in the area of your activity (PoS$_2$), are selected automatically.

\section{Issues and Challenges}

Despite the use of IoE and wearable sensors, a number of challenges including privacy, security, scalability, and quality of service might impact the use of the proposed solution. We briefly highlight the issues and their probable solution. 

\begin{itemize}
    \item \textit{Challenge related to privacy}: Data privacy is of major concern for the CORFIE framework as the data gathered from wearable and mobile sensors contain personal information and an individual would not risk the data privacy even if it helps to protect him/her from the contagious virus. Furthermore, information theft due to the privacy breach can be used for illegal benefits. The privacy issue has been highlighted by some legal frameworks such as the health insurance portability and accountability act (HIPAA) \cite{Nosowsky2006} and general data protection regulation (GDPR) \cite{Goddard2017}. 
    \item \textit{Potential solutions and research directions for privacy concerns}: To deal with privacy issues, one of the solutions is to move the analytical processes to the middleware or edge devices having light-weight operations for critical or highly sensitive data while transmitting only the decision label to the server for further action. Another solution to the privacy issue is the use of software defined privacy \cite{Kemmer2016}, privacy by design \cite{Semantha2020}, Federated learning \cite{Hao2020}, and other solutions in compliance with the privacy life-cycle. 
    \item \textit{Challenge related to security}: The security issues have garnered a lot of attention from researchers recently due to the versatility of attacks and fast pace modifications. The data collected from wearable sensors are quite sensitive due to the high-risk involved with the decision analysis part. Such attacks could manipulate the data for the change of decision at the analysis stage which might result in actions involving risks or negative impact. The security issues are also directly related to scalability suggesting that the security gets vulnerable with the increase of devices. 
    \item \textit{Potential solutions and research directions for security concerns}: One of the possible solutions is similar to the privacy concern, i.e. to move essential services to the edge devices or middleware for reducing the communication flow of the data to the server. The CORFIE framework can be fused with a secure REST approach or lightweight anonymous authentication protocol to increase network security. Conventional authentication mechanisms such as IDs and passwords can be used for securing the data. However, for more efficient approaches the use of machine learning and distributed services such as Blockchain can be leveraged to secure the data, accordingly. 
    \item \textit{Challenge related to scalability and quality of services}: As the number of wearable sensors and devices is increasing drastically, it poses scalability and quality of service issues to a great extent. Similar to the devices and with the emergence of IoE, the number of users and other service requesting entities have also increased manifold. The service provision to all the users and devices lead to network congestion problem which could probably delay the decision outcome of CORFIE framework. In this regard, the framework and the network service provider both need to support scalability in order to facilitate hard- and soft-real time systems, especially in these desperate times. With reference to the increase in devices, it requires the framework to handle the data generation and storage process. 
    \item \textit{Potential solutions and research directions for scalability and quality of service concerns}: Distributed data storage services could be used which include Cassandra, MongoDB, and Apache HBase. Keeping in mind the CORFIE framework, the middleware could use “s3cmd utility” to send the acquired data to Amazon storage in a distributed manner which then will be permanently stored to the online server. For network scalability, many methods based on machine learning and optimization have been proposed for active user detection in 5G/6G networks. This benefits the CORFIE framework in two ways. The first is the reduction of communication between devices that are not involved in the desired service being selected and the second is the grant-free access to the base station in an efficient way. Moreover, virtual software network functions and network slicing approaches can also be used to reduce the network congestion problem, respectively.   
    \item \textit{Challenge related to society}: Societal issues due to the continuously revolving conspiracy theories and technology acceptance have been there for most of the advancements that have happened in the last decades. Recently, the link of COVID-19 and 5G technology has made a lot of headlines which not only impacts the technology acceptance but also affects the intrinsic issues of a particular new technology and advancements in the specific directions. 
    \item \textit{Potential solutions and research directions for societal issues}: As mentioned earlier regarding the acceptance of the COVID-19 vaccination, the employers and the governments could be taken in confidence in order to provide technical literacy regarding the breakthrough or advancements. The collaboration with governments and employers could reduce the impact of conspiracies and help the people to accept the technology which is mainly proposed to assist people with their everyday activities as well as facilitate them in protecting from the contagious virus.
    \item \textit{Challenges related to legal implications}: Whenever the collection of data is involved especially if the same data is used to track your activities without your consent, the legal issues will pave their way. Furthermore, the collection of such data without consent have legal implications. This problem is also related to technology acceptance and so does its solution.
    \item \textit{Potential solutions and research directions for legal implications}: Similar to the participants who are willing to volunteer for vaccines, employers, and government organizations, in collaboration, can call for participants who are willing to volunteer for providing their data regarding contact tracing. This procedure could also involve standardization and regulatory bodies to provide guidelines regarding the collection and use of such data. It will not only reduce the legal implications but will help individuals (who have faith in their employers and governments) to understand the importance of such systems. 
\end{itemize}

There are several other constraints, challenges, and research directions that include connectivity issues, battery consumption of the middleware and wearable sensors, memory profiling, and others but those have not been touched upon in this work. 

\section{Conclusion}
\textit{The future is unpredictable but so do the events that lead to that future}. It has been established that vaccines for coronavirus and in case of future pandemics will either take time to reach out in every corner of the world or need continuous modifications/alterations due to the mutations and variants. In the latter case, the life-cycle of vaccine developments needs to be revised along with the time to get approval from the drug regulatory authority. It has also been learned in previous months while dealing with pandemics that if one occurs in the future, the solution of vaccines will take at least 2 – 3 years for reaching to an individual (depending on the several medical and technical constraints) and that the good practices, immunity-boosting, personal hygiene, healthcare monitoring, and contact tracing might help in slowing down the transmission of infection. The aforementioned good practices can be brought to realization with macro solutions based on wearable/mobile sensors and integrated technologies.
In this article, we emphasize the importance of using integrated technologies to help in dealing with the COVID-19 and future pandemics till the vaccine shows up at the doorstep. We presented hypothetical frameworks for physical healthcare monitoring, personal hygiene and immunity-boosting, mental healthcare, and contract tracing applications. Furthermore, we presented COvid-19 Resistance Framework using the Internet of Everything (CORFIE) which combines the aforementioned modules into a single framework. We briefly defined the technical details and provided a summary of potential challenges along with the probable solutions and research directions to make CORFIE a realization. We assume that CORFIE in general can prevent individuals from contracting novel coronavirus while adopting good practices. We believe that CORFIE will not only highlight the importance of using wearable/mobile sensors for individual prevention but also will help in making statuary bodies, governments, and industries redirect some part of their funding to the proposed initiative. 
We believe that the proposed work will also shed light on potential future works that could be branched off from the CORFIE architecture. We intend to develop an Android app based on the CORFIE architectural attributes and to show the potential benefits while avoiding any legal implications in terms of data collection and analysis. We also intend to conduct an agent-based simulation that could help understand the economic benefits if the CORFIE architecture is used at a large scale.    

\bibliographystyle{IEEEtran}
\bibliography{IEEEabrv,ref}

\begin{thebibliography}{10}
\providecommand{\url}[1]{#1}
\csname url@samestyle\endcsname
\providecommand{\newblock}{\relax}
\providecommand{\bibinfo}[2]{#2}
\providecommand{\BIBentrySTDinterwordspacing}{\spaceskip=0pt\relax}
\providecommand{\BIBentryALTinterwordstretchfactor}{4}
\providecommand{\BIBentryALTinterwordspacing}{\spaceskip=\fontdimen2\font plus
\BIBentryALTinterwordstretchfactor\fontdimen3\font minus
  \fontdimen4\font\relax}
\providecommand{\BIBforeignlanguage}[2]{{%
\expandafter\ifx\csname l@#1\endcsname\relax
\typeout{** WARNING: IEEEtran.bst: No hyphenation pattern has been}%
\typeout{** loaded for the language `#1'. Using the pattern for}%
\typeout{** the default language instead.}%
\else
\language=\csname l@#1\endcsname
\fi
#2}}
\providecommand{\BIBdecl}{\relax}
\BIBdecl

\bibitem{meter}
\BIBentryALTinterwordspacing
``{Coronavirus Update (Live)},'' 2020. [Online]. Available:
  \url{https://www.worldometers.info/coronavirus/}
\BIBentrySTDinterwordspacing

\bibitem{Dev2020}
\BIBentryALTinterwordspacing
K.~Dev, S.~A. Khowaja, A.~S. Bist, V.~Saini, and S.~Bhatia, ``{Triage of
  Potential COVID-19 Patients from Chest X-ray Images using Hierarchical
  Convolutional Networks},'' nov 2020. [Online]. Available:
  \url{http://arxiv.org/abs/2011.00618}
\BIBentrySTDinterwordspacing

\bibitem{Dickinson2020}
\BIBentryALTinterwordspacing
D.~Dickinson, ``{Young people ‘not invincible' in COVID-19 pandemic: WHO
  chief},'' 2020. [Online]. Available:
  \url{https://news.un.org/en/story/2020/07/1069301}
\BIBentrySTDinterwordspacing

\bibitem{geno}
\BIBentryALTinterwordspacing
``{COG UK News and Updates},'' 2020. [Online]. Available:
  \url{https://www.cogconsortium.uk/news/}
\BIBentrySTDinterwordspacing

\bibitem{Wise2020}
\BIBentryALTinterwordspacing
J.~Wise, ``{Covid-19: New coronavirus variant is identified in UK},''
  \emph{BMJ}, p. m4857, dec 2020. [Online]. Available:
  \url{https://www.bmj.com/lookup/doi/10.1136/bmj.m4857}
\BIBentrySTDinterwordspacing

\bibitem{TheLancetMicrobe2020}
\BIBentryALTinterwordspacing
{The Lancet Microbe}, ``{COVID-19 vaccines: the pandemic will not end
  overnight},'' \emph{The Lancet Microbe}, dec 2020. [Online]. Available:
  \url{https://linkinghub.elsevier.com/retrieve/pii/S2666524720302263}
\BIBentrySTDinterwordspacing

\bibitem{col}
\BIBentryALTinterwordspacing
``{COVID-19 Vaccine frequently asked questions},'' 2020. [Online]. Available:
  \url{https://covid19.colorado.gov/vaccine-faq}
\BIBentrySTDinterwordspacing

\bibitem{voysey2020safety}
M.~Voysey, S.~A.~C. Clemens, S.~A. Madhi, L.~Y. Weckx, P.~M. Folegatti, P.~K.
  Aley, B.~Angus, V.~L. Baillie, S.~L. Barnabas, Q.~E. Bhorat \emph{et~al.},
  ``Safety and efficacy of the chadox1 ncov-19 vaccine (azd1222) against
  sars-cov-2: an interim analysis of four randomised controlled trials in
  brazil, south africa, and the uk,'' \emph{The Lancet}, 2020.

\bibitem{polack2020safety}
F.~P. Polack, S.~J. Thomas, N.~Kitchin, J.~Absalon, A.~Gurtman, S.~Lockhart,
  J.~L. Perez, G.~P{\'e}rez~Marc, E.~D. Moreira, C.~Zerbini \emph{et~al.},
  ``Safety and efficacy of the bnt162b2 mrna covid-19 vaccine,'' \emph{New
  England Journal of Medicine}, 2020.

\bibitem{Cyranoski2020}
\BIBentryALTinterwordspacing
D.~Cyranoski, ``{Arab nations first to approve Chinese COVID vaccine —
  despite lack of public data},'' 2020. [Online]. Available:
  \url{https://www.nature.com/articles/d41586-020-03563-z}
\BIBentrySTDinterwordspacing

\bibitem{Schmidt2020}
\BIBentryALTinterwordspacing
C.~Schmidt, ``{Fauci Explains How to End the COVID Pandemic},'' 2020. [Online].
  Available:
  \url{https://www.scientificamerican.com/article/fauci-explains-how-to-end-the-covid-pandemic1/}
\BIBentrySTDinterwordspacing

\bibitem{Mccoy2020}
\BIBentryALTinterwordspacing
J.~Mccoy, ``{The COVID-19 Vaccine Is Rolling Out Across Colorado. But When Will
  the Local Epidemic Actually Be Over?}'' 2020. [Online]. Available:
  \url{https://www.5280.com/2020/12/the-covid-19-vaccine-is-rolling-out-across-colorado-but-when-will-
  \\ -local-epidemic-actually-be-over/}
\BIBentrySTDinterwordspacing

\bibitem{Lazarus2020}
\BIBentryALTinterwordspacing
J.~V. Lazarus, S.~C. Ratzan, A.~Palayew, L.~O. Gostin, H.~J. Larson, K.~Rabin,
  S.~Kimball, and A.~El-Mohandes, ``{A global survey of potential acceptance of
  a COVID-19 vaccine},'' \emph{Nature Medicine}, pp. 1--9, oct 2020. [Online].
  Available: \url{http://www.nature.com/articles/s41591-020-1124-9}
\BIBentrySTDinterwordspacing

\bibitem{Dai2020}
\BIBentryALTinterwordspacing
B.~Dai, E.~Larnyo, E.~A. Tetteh, A.~K. Aboagye, and A.-A.~I. Musah, ``{Factors
  Affecting Caregivers' Acceptance of the Use of Wearable Devices by Patients
  With Dementia: An Extension of the Unified Theory of Acceptance and Use of
  Technology Model},'' \emph{American Journal of Alzheimer's Disease {\&} Other
  Dementias{\textregistered}}, vol.~35, p. 153331751988349, jan 2020. [Online].
  Available: \url{http://journals.sagepub.com/doi/10.1177/1533317519883493}
\BIBentrySTDinterwordspacing

\bibitem{Li2019}
\BIBentryALTinterwordspacing
J.~Li, Q.~Ma, A.~H. Chan, and S.~Man, ``{Health monitoring through wearable
  technologies for older adults: Smart wearables acceptance model},''
  \emph{Applied Ergonomics}, vol.~75, pp. 162--169, feb 2019. [Online].
  Available:
  \url{https://linkinghub.elsevier.com/retrieve/pii/S0003687018305167}
\BIBentrySTDinterwordspacing

\bibitem{Wang2020}
\BIBentryALTinterwordspacing
H.~Wang, D.~Tao, N.~Yu, and X.~Qu, ``{Understanding consumer acceptance of
  healthcare wearable devices: An integrated model of UTAUT and TTF},''
  \emph{International Journal of Medical Informatics}, vol. 139, p. 104156, jul
  2020. [Online]. Available:
  \url{https://linkinghub.elsevier.com/retrieve/pii/S1386505619311438}
\BIBentrySTDinterwordspacing

\bibitem{Dutot2019}
\BIBentryALTinterwordspacing
V.~Dutot, V.~Bhatiasevi, and N.~Bellallahom, ``{Applying the technology
  acceptance model in a three-countries study of smartwatch adoption},''
  \emph{The Journal of High Technology Management Research}, vol.~30, no.~1,
  pp. 1--14, may 2019. [Online]. Available:
  \url{https://linkinghub.elsevier.com/retrieve/pii/S1047831019300033}
\BIBentrySTDinterwordspacing

\bibitem{Khowaja2018}
\BIBentryALTinterwordspacing
S.~A. Khowaja, A.~G. Prabono, F.~Setiawan, B.~N. Yahya, and S.-L. Lee,
  ``{Contextual activity based Healthcare Internet of Things, Services, and
  People (HIoTSP): An architectural framework for healthcare monitoring using
  wearable sensors},'' \emph{Computer Networks}, vol. 145, pp. 190--206, nov
  2018. [Online]. Available:
  \url{https://linkinghub.elsevier.com/retrieve/pii/S1389128618308594}
\BIBentrySTDinterwordspacing

\bibitem{Khowaja2017}
\BIBentryALTinterwordspacing
S.~A. Khowaja, B.~N. Yahya, and S.-L. Lee, ``{Hierarchical classification
  method based on selective learning of slacked hierarchy for activity
  recognition systems},'' \emph{Expert Systems with Applications}, vol.~88, pp.
  165--177, dec 2017. [Online]. Available:
  \url{https://linkinghub.elsevier.com/retrieve/pii/S0957417417304645}
\BIBentrySTDinterwordspacing

\bibitem{Khowaja2016}
S.~A. Khowaja, F.~Setiawan, A.~G. Prabono, B.~N. Yahya, and S.-L. Lee, ``{An
  Effective Threshold Based Measurement Technique for Fall Detection Using
  Smart Devices},'' \emph{International Journal of Industrial Engineering:
  Theory, Applications, and Practice}, vol.~23, no.~5, 2016.

\bibitem{Khowaja2020a}
\BIBentryALTinterwordspacing
S.~A. Khowaja, B.~N. Yahya, and S.-L. Lee, ``{CAPHAR: context-aware
  personalized human activity recognition using associative learning in smart
  environments},'' \emph{Human-centric Computing and Information Sciences},
  vol.~10, no.~1, p.~35, dec 2020. [Online]. Available:
  \url{https://hcis-journal.springeropen.com/articles/10.1186/s13673-020-00240-y}
\BIBentrySTDinterwordspacing

\bibitem{Chaudhuri2020}
\BIBentryALTinterwordspacing
S.~Chaudhuri, S.~Basu, P.~Kabi, V.~R. Unni, and A.~Saha, ``{Modeling the role
  of respiratory droplets in Covid-19 type pandemics},'' \emph{Physics of
  Fluids}, vol.~32, no.~6, p. 063309, jun 2020. [Online]. Available:
  \url{http://aip.scitation.org/doi/10.1063/5.0015984}
\BIBentrySTDinterwordspacing

\bibitem{Fontes2020}
\BIBentryALTinterwordspacing
D.~Fontes, J.~Reyes, K.~Ahmed, and M.~Kinzel, ``{A study of fluid dynamics and
  human physiology factors driving droplet dispersion from a human sneeze},''
  \emph{Physics of Fluids}, vol.~32, no.~11, p. 111904, nov 2020. [Online].
  Available: \url{http://aip.scitation.org/doi/10.1063/5.0032006}
\BIBentrySTDinterwordspacing

\bibitem{Khoramipour2020}
\BIBentryALTinterwordspacing
K.~Khoramipour, A.~Basereh, A.~A. Hekmatikar, L.~Castell, R.~T. Ruhee, and
  K.~Suzuki, ``{Physical activity and nutrition guidelines to help with the
  fight against COVID-19},'' \emph{Journal of Sports Sciences}, pp. 1--7, aug
  2020. [Online]. Available:
  \url{https://www.tandfonline.com/doi/full/10.1080/02640414.2020.1807089}
\BIBentrySTDinterwordspacing

\bibitem{Damiot2020}
\BIBentryALTinterwordspacing
A.~Damiot, A.~J. Pinto, J.~E. Turner, and B.~Gualano, ``{Immunological
  Implications of Physical Inactivity among Older Adults during the COVID-19
  Pandemic},'' \emph{Gerontology}, vol.~66, no.~5, pp. 431--438, 2020.
  [Online]. Available: \url{https://www.karger.com/Article/FullText/509216}
\BIBentrySTDinterwordspacing

\bibitem{Vorvick2020}
\BIBentryALTinterwordspacing
L.~J. Vorvick and D.~Zieve, ``{Exercise and immunity: Medline Plus},'' 2020.
  [Online]. Available: \url{https://medlineplus.gov/ency/article/007165.htm}
\BIBentrySTDinterwordspacing

\bibitem{Fofana2020}
\BIBentryALTinterwordspacing
N.~K. Fofana, F.~Latif, S.~Sarfraz, Bilal, M.~F. Bashir, and B.~Komal, ``{Fear
  and agony of the pandemic leading to stress and mental illness: An emerging
  crisis in the novel coronavirus (COVID-19) outbreak},'' \emph{Psychiatry
  Research}, vol. 291, p. 113230, sep 2020. [Online]. Available:
  \url{https://linkinghub.elsevier.com/retrieve/pii/S0165178120310970}
\BIBentrySTDinterwordspacing

\bibitem{Menzies2020}
\BIBentryALTinterwordspacing
R.~E. Menzies and R.~G. Menzies, ``{Death anxiety in the time of COVID-19:
  theoretical explanations and clinical implications},'' \emph{The Cognitive
  Behaviour Therapist}, vol.~13, p. e19, jun 2020. [Online]. Available:
  \url{https://www.cambridge.org/core/product/identifier/S1754470X20000215/type/journal{\_}article}
\BIBentrySTDinterwordspacing

\bibitem{Madigan2020}
\BIBentryALTinterwordspacing
S.~Madigan, N.~Racine, J.~E. Cooke, and D.~J. Korczak, ``{COVID-19 and
  telemental health: Benefits, challenges, and future directions.}''
  \emph{Canadian Psychology/Psychologie canadienne}, oct 2020. [Online].
  Available: \url{http://doi.apa.org/getdoi.cfm?doi=10.1037/cap0000259}
\BIBentrySTDinterwordspacing

\bibitem{NationalCenterforImmunizationandRespiratoryDiseasesNCIRD2020}
\BIBentryALTinterwordspacing
D.~o. V.~D. {National Center for Immunization and Respiratory Diseases
  (NCIRD)}, ``{Coping with Stress},'' 2020. [Online]. Available:
  \url{https://www.cdc.gov/coronavirus/2019-ncov/daily-life-coping/managing-stress-anxiety.html}
\BIBentrySTDinterwordspacing

\bibitem{Sher2020}
\BIBentryALTinterwordspacing
L.~Sher, ``{The impact of the COVID-19 pandemic on suicide rates},'' \emph{QJM:
  An International Journal of Medicine}, vol. 113, no.~10, pp. 707--712, oct
  2020. [Online]. Available:
  \url{https://academic.oup.com/qjmed/article/113/10/707/5857612}
\BIBentrySTDinterwordspacing

\bibitem{Ettman2020}
\BIBentryALTinterwordspacing
C.~K. Ettman, S.~M. Abdalla, G.~H. Cohen, L.~Sampson, P.~M. Vivier, and
  S.~Galea, ``{Prevalence of Depression Symptoms in US Adults Before and During
  the COVID-19 Pandemic},'' \emph{JAMA Network Open}, vol.~3, no.~9, p.
  e2019686, sep 2020. [Online]. Available:
  \url{https://jamanetwork.com/journals/jamanetworkopen/fullarticle/2770146}
\BIBentrySTDinterwordspacing

\bibitem{McGinty2020}
\BIBentryALTinterwordspacing
E.~E. McGinty, R.~Presskreischer, H.~Han, and C.~L. Barry, ``{Psychological
  Distress and Loneliness Reported by US Adults in 2018 and April 2020},''
  \emph{JAMA}, vol. 324, no.~1, p.~93, jul 2020. [Online]. Available:
  \url{https://jamanetwork.com/journals/jama/fullarticle/2766941}
\BIBentrySTDinterwordspacing

\bibitem{Setiawan2018}
\BIBentryALTinterwordspacing
F.~Setiawan, S.~A. Khowaja, A.~G. Prabono, B.~N. Yahya, and S.-L. Lee, ``{A
  Framework for Real Time Emotion Recognition Based on Human ANS Using
  Pervasive Device},'' in \emph{2018 IEEE 42nd Annual Computer Software and
  Applications Conference (COMPSAC)}.\hskip 1em plus 0.5em minus 0.4em\relax
  IEEE, jul 2018, pp. 805--806. [Online]. Available:
  \url{https://ieeexplore.ieee.org/document/8377754/}
\BIBentrySTDinterwordspacing

\bibitem{Khowaja2020}
\BIBentryALTinterwordspacing
S.~A. Khowaja, A.~G. Prabono, F.~Setiawan, B.~N. Yahya, and S.-L. Lee,
  ``{Toward soft real-time stress detection using wrist-worn devices for human
  workspaces},'' \emph{Soft Computing}, sep 2020. [Online]. Available:
  \url{http://link.springer.com/10.1007/s00500-020-05338-0}
\BIBentrySTDinterwordspacing

\bibitem{Chowdhury2020}
\BIBentryALTinterwordspacing
M.~J.~M. Chowdhury, M.~S. Ferdous, K.~Biswas, N.~Chowdhury, and
  V.~Muthukkumarasamy, ``{COVID-19 Contact Tracing: Challenges and Future
  Directions},'' \emph{IEEE Access}, pp. 1--1, 2020. [Online]. Available:
  \url{https://ieeexplore.ieee.org/document/9252092/}
\BIBentrySTDinterwordspacing

\bibitem{Ahmed2020}
\BIBentryALTinterwordspacing
N.~Ahmed, R.~A. Michelin, W.~Xue, S.~Ruj, R.~Malaney, S.~S. Kanhere,
  A.~Seneviratne, W.~Hu, H.~Janicke, and S.~K. Jha, ``{A Survey of COVID-19
  Contact Tracing Apps},'' \emph{IEEE Access}, vol.~8, pp. 134\,577--134\,601,
  2020. [Online]. Available:
  \url{https://ieeexplore.ieee.org/document/9144194/}
\BIBentrySTDinterwordspacing

\bibitem{Kretzschmar2020}
\BIBentryALTinterwordspacing
M.~E. Kretzschmar, G.~Rozhnova, M.~C.~J. Bootsma, M.~van Boven, J.~H. H.~M.
  van~de Wijgert, and M.~J.~M. Bonten, ``{Impact of delays on effectiveness of
  contact tracing strategies for COVID-19: a modelling study},'' \emph{The
  Lancet Public Health}, vol.~5, no.~8, pp. e452--e459, aug 2020. [Online].
  Available:
  \url{https://linkinghub.elsevier.com/retrieve/pii/S2468266720301572}
\BIBentrySTDinterwordspacing

\bibitem{Nosowsky2006}
\BIBentryALTinterwordspacing
R.~Nosowsky and T.~J. Giordano, ``{The Health Insurance Portability and
  Accountability Act of 1996 (HIPAA) Privacy Rule: Implications for Clinical
  Research},'' \emph{Annual Review of Medicine}, vol.~57, no.~1, pp. 575--590,
  feb 2006. [Online]. Available:
  \url{http://www.annualreviews.org/doi/10.1146/annurev.med.57.121304.131257}
\BIBentrySTDinterwordspacing

\bibitem{Goddard2017}
\BIBentryALTinterwordspacing
M.~Goddard, ``{The EU General Data Protection Regulation (GDPR): European
  Regulation that has a Global Impact},'' \emph{International Journal of Market
  Research}, vol.~59, no.~6, pp. 703--705, nov 2017. [Online]. Available:
  \url{http://journals.sagepub.com/doi/10.2501/IJMR-2017-050}
\BIBentrySTDinterwordspacing

\bibitem{Kemmer2016}
\BIBentryALTinterwordspacing
F.~Kemmer, C.~Reich, M.~Knahl, and N.~Clarke, ``{Software Defined Privacy},''
  in \emph{IEEE International Conference on Cloud Engineering Workshop
  (IC2EW)}.\hskip 1em plus 0.5em minus 0.4em\relax IEEE, apr 2016, pp. 25--29.
  [Online]. Available: \url{http://ieeexplore.ieee.org/document/7527810/}
\BIBentrySTDinterwordspacing

\bibitem{Semantha2020}
\BIBentryALTinterwordspacing
F.~H. Semantha, S.~Azam, K.~C. Yeo, and B.~Shanmugam, ``{A Systematic
  Literature Review on Privacy by Design in the Healthcare Sector},''
  \emph{Electronics}, vol.~9, no.~3, p. 452, mar 2020. [Online]. Available:
  \url{https://www.mdpi.com/2079-9292/9/3/452}
\BIBentrySTDinterwordspacing

\bibitem{Hao2020}
\BIBentryALTinterwordspacing
M.~Hao, H.~Li, X.~Luo, G.~Xu, H.~Yang, and S.~Liu, ``{Efficient and
  Privacy-Enhanced Federated Learning for Industrial Artificial
  Intelligence},'' \emph{IEEE Transactions on Industrial Informatics}, vol.~16,
  no.~10, pp. 6532--6542, oct 2020. [Online]. Available:
  \url{https://ieeexplore.ieee.org/document/8859260/}
\BIBentrySTDinterwordspacing

\end{thebibliography}
\begin{IEEEbiographynophoto}{Dr. Sunder Ali Khowaja} is an Assistant Professor at Department of Telecommunication, Faculty of Engineering and Technology, University of Sindh, Jamshoro, Pakistan. He has worked with Comstar ISA Ltd. and New Horizon IT Ltd. as RF/VSAT Engineer and Network Engineer from 2008-2011.\\
He was awarded the Ph.D. degree in Industrial and Information Systems Engineering from Hankuk University of Foreign Studies, South Korea funded by Foreign Students Scholarship. He has secured M.E. degree in Communication Systems and Networks; and B.E. degree in Telecommunication from Mehran University of Engineering and Technology, Pakistan. He is serving as a reviewer for many reputed journals including IEEE Access, IET Image Processing, IET Electronic Letters, IET Signal Processing, IET Wireless Sensor Systems, International Journal of Imaging Systems and Technology, Artificial Intelligence Review, Computational and Mathematical Methods in Medicine, Computers in Human Behavior, Mathematical Problems in Engineering, Multimedia Systems, and Neural Processing Letters. His research interests include Data Analytics and Machine Learning for Ambient Intelligence, Image Analysis and Computer Vision applications.
\end{IEEEbiographynophoto}
\vskip -2\baselineskip plus -1fil
\vspace{-0.25cm}
\begin{IEEEbiographynophoto}{Parus Khuwaja } is pursuing her Ph.D. degree in financial analytics from University of Sindh, Jamshoro. She is currently working as an Assistant Professor at Institute of Business Administration, University of Sindh, Jamshoro. Her interests include Data analytics, Machine learning for Ambient Intelligence, Stock Portfolios, and Financial securities.
\end{IEEEbiographynophoto}
\vskip -2\baselineskip plus -1fil
\vspace{-0.25cm}
\begin{IEEEbiographynophoto}{Kapal Dev} is Senior Researcher at Nimbus Research Centre, Munster Technological University, Ireland. Previously, he was a Postdoctoral Research Fellow with the CONNECT Centre, School of Computer Science and Statistics, Trinity College Dublin (TCD). Previously, he worked as 5G Junior Consultant and Engineer at Altran Italia S.p.A, Milan, Italy on 5G use cases. He worked as Lecturer at Indus university, Karachi back in 2014. He is also working for OCEANS Network as Head of Projects to manage OCEANS project processes and functions to improve efficiency, consistency and best practice integration. \\
He was awarded the PhD degree by Politecnico di Milano, Italy under the prestigious fellowship of Erasmus Mundus funded by European Commission. His education Profile revolves over ICT background i.e. Electronics (B.E and M.E), Telecommunication Engineering (PhD) and Post-doc (Fusion of 5G and Blockchain). His research interests include Blockchain, 6G Networks and Artificial Intelligence. He is very active in leading (as Principle Investigator) Erasmus + International Credit Mobility (ICM) and Capacity Building for Higher Education and H2020 Co-Fund projects. Received a few Million Euros funding and few are in review as well as in the writing phase. \\
He is evaluator of MSCA Co-Fund schemes, Elsevier Book proposals and top scientific journals and conferences including IEEE TII, IEEE TITS, IEEE TNSE, IEEE IoT, IEEE JBHI, FGCS, COMNET, TETT, IEEE VTC, WF-IoT. TPC member of IEEE BCA 2020 in conjunction with AICCSA 2020, ICBC 2021, DICG Co-located with Middleware 2020 and FTNCT 2020. He is also serving as Associate Editor in Wireless Networks, IET Quantum Communication and Review Editor in Frontiers in Communications and Networks. He is also serving as Guest Editor (GE) in COMCOM (I.F: 2.8), GE in COMNET (I.F 3.11), Lead chair in one of CCNC 2021 workshops, and editing a book for CRC press.  He  is  a  member  of  the  ACM, IEEE,  and  actively involved  with  various  working  groups  and committees  of  IEEE  and  ACM  related  to 5G and beyond,  Blockchain  and Artificial Intelligence. \\
\end{IEEEbiographynophoto}
\end{document}